\documentclass[hyper]{JHEP} 

\usepackage{epsfig}

\newcommand\fverb{\setbox\pippobox=\hbox\bgroup\verb}

\newcommand\fverbdo{\egroup\medskip\noindent%

            \fbox{\unhbox\pippobox}\ }

\newcommand\fverbit{\egroup\item[\fbox{\unhbox\pippobox}]}

\newbox\pippobox

\title{Dynamics of D1-brane in I-brane Background}

\author{J. Kluso\v{n}
 \footnote{On leave from Masaryk University, Brno}\\
Dipartimento di Fisica,\\
Universita' di Roma \& Sezione di Roma 2, Tor Vergata \\
Via della, Ricerca, Scientifica, 1 00133  Roma   ITALY\\
E-mail:
\email{Josef.Kluson@roma2.infn.it}}

\preprint{\hepth{0510243}}

 \abstract{This paper is devoted to the study of the
 effective field theory description
 of the probe D1-brane in the background of the system
 of two stacks of fivebranes in type IIB theory that
 intersect on  $R^{1,1}$. We study the properties
 of the Dirac-Born-Infeld action for D1-brane
 moving in this background. We will argue that
 this action is invariant under
  an additional symmetry in the near horizon
limit and that this new symmetry is
closely related to the enhanced
symmetry of the I-brane background
considered recently in
[hep-th/0508025]. We also solve
explicitly the equation of motion of
D1-brane in the near horizon limit.}


\def\bZ{\mathbf{Z}}
\def\bz{\mathbf{z}}
\def\by{\mathbf{y}}

\def\bA{\mathbf{A}}

\def\mK{\mathcal{K}}
\def\mH{\mathcal{H}}
\def\bAi{\left(\mathbf{A}^{-1}\right)}
\def\bZ{\mathbf{Z}}
\def\bY{\mathbf{Y}}
\def\mL{\mathcal{L}}

\def\tR{S}
\begin{document}
\section{Introduction and Summary}\label{first}
One of the most outstanding problems in
string theory is  the problem of string
dynamics in the time dependent
background. On the other hand there was
great recent progress pioneered by work
of A. Sen
\cite{Sen:2002nu,Sen:2002in,Sen:2002an,Sen:2003tm}
that showed that the rolling tachyon
could be used to describe an decay of
an unstable D-brane. It is very nice
that Dirac-Born-Infeld (DBI)-like
tachyon effective  action
\cite{Sen:1999md,Bergshoeff:2000dq,Garousi:2000tr,Kluson:2000iy}
 can capture most of
the aspects of the time dynamics of the
tachyon.

The problem of the time dependence of
the string theory was also analyzed
from different point of view in the
paper
\cite{Kutasov:2005rr}\footnote{Some
related works are
\cite{Nakayama:2005pk,Kluson:2005ab,
Kluson:2005dr,Huang:2005hy,Kluson:2005mr,Thomas:2005fw,Kluson:2005jr,
Huang:2005rd,Chen:2005wm,
Kluson:2005qx,Nakayama:2004ge,Chen:2004vw,
Thomas:2004cd,Bak:2004tp,
Kluson:2004yk,Kluson:2004xc,
Kutasov:2004ct,Sahakyan:2004cq,
Ghodsi:2004wn,Panigrahi:2004qr,Saremi:2004yd}.}
 where the time dependent motion of
probe Dp-brane in the background of $N$
NS5-branes was considered. Among many
interesting results derived there was
an observation that the radial motion
of such a probe Dp-brane can be mapped
to the time depend tachyon like
condensation. It is then very tempting
to conjecture that tachyon could have
geometrical origin.
  Kutasov's paper is
also an example how probe Dp-brane
dynamics could be useful for the study
of properties of  given background. In
fact the probe D-brane analysis was
very successful in the study of matrix
theory (See
\cite{Taylor:1999qk,Taylor:1997dy} for
reviews and extensive list of
references.) and AdS/CFT correspondence
(See for example
\cite{Aharony:1999ti}.) For that reason
it seems to be natural to  analyze the
dynamics of  the probe D-brane  in all
interesting backgrounds. One such an
example of a nontrivial string theory
 background was recently considered  in
a remarkable paper by N. Itzaki, D.
Kutasov and N. Seiberg
\cite{Itzhaki:2005tu}. This background
consists two stacks of fivebranes in
type IIB theory that intersect on
$R^{1,1}$. Very careful analysis
presented there discovered many
interesting results. For example, it
was shown that there is an enhancement
of the symmetry in the  near horizon
geometry of given background  that in
some sense mixes the world volume
coordinate of I-brane with the
transverse ones \footnote{This point
was also confirmed in
\cite{Lin:2005nh}.}. This
 property was then  analyzed in
\cite{Itzhaki:2005tu} from the point of
view of the effective field theory
living on the intersecting NS5-branes
and also from the perspective of the
closed string moving in this
background. Unfortunately there is not
enough place to review all  results
derived in this paper and we recommend
the original paper
\cite{Itzhaki:2005tu} for more details
with profound physical interpretation.

The goal of this paper is much modest.
We will study  the classical dynamics
of the probe D1-brane in the I-brane
background, where one can hope,
according to arguments given above,
that this dynamics could reflect some
of the nontrivial predictions given in
paper \cite{Itzhaki:2005tu}. To do this
we will carefully examine the conserved
charges for D1-brane moving in given
background. We will see that in the
near horizon limit of the I-brane
geometry one can find on the world
volume of D1-brane a new
transformation. It turns out that
this-scaling like transformation-is
symmetry of the action  in case of the
pure time dependent dynamics on the
world volume of D1-brane and also in
case when all world volume scalar modes
are spatial dependent only
\footnote{The case of spatial dependent
scalar fields will be analyzed in
separate publication.}. Then with the
help of the new conserved charge we
will be able to solve  the equations of
motion explicitly. Following
\cite{Nakayama:2005pk,Sahakyan:2004cq}
we  could expect that this time
evolution of the probe D1-brane in
I-brane background has holographic
description in a  dual $2+1$
dimensional Little String Theory
\cite{Itzhaki:2005tu}.

Despite of the fact that
 the analysis given in this
paper cannot much to say to the
holographic formulation of I-dynamics
we can still hope that it  could be
helpful for the study of the
interesting properties of I-brane
background. For example, an emergence
of the scaling symmetry in the near
horizon region is related to the
enhancement of the world volume
symmetry of I-brane from $(1,1)$ to
$(2,1)$. We also hope that the
conformal field theory description of
D-brane dynamics in I-brane background
will be determined and the solutions
found in this paper
 could arise as a classical limit of the
exact boundary state analysis.  On the
other hand we still believe that
 the DBI analysis of the probe could
be extended in several directions.
First of all, we are going to study the
spatial dependent solutions of the
D1-brane in the I-brane background. We
can also study the dynamics of D1-brane
probe in the background where some of
the NS5-branes are not coincident.
Another interesting problem seems to be
the analysis of   probe Dp-brane
dynamics in the background of  $1+1$
dimensional intersection of two
orthogonal stacks of NS5-branes
 in the type IIA theory or in the
 background of $1+1$ dimensional
 intersection of two orthogonal stack of
 D5-branes in type IIB theory.

The organization of the paper is as
follows. In the next section
(\ref{second}) we will review the
properties of the DBI action in general
background. We will also study the
symmetry properties of given action and
we will outline the general
prescription  how to obtain charges
that either are conserved or obey the
equation where on the left hand side is
a divergence of given current while the
right hand side contains an  anomaly
term that is a result of the explicit
breaking of given symmetry in given
action. In section (\ref{third}) we
will introduce the Lagrangian density
for probe D1-brane in I-brane
background and determine some conserved
quantities that follow from the
symmetry of D1-brane action in
$I$-brane background.
 In section
(\ref{fourth}) we will solve the
equation of motion for D1-brane in the
near horizon limit where all modes on
the world volume of D1-brane are
homogeneous. We find explicit time
dependence of the radial modes that
describe the radial position of
D1-brane. We will also find the
relation between the dilatation
symmetry that emerges on the world
volume of D1-brane and the new symmetry
that was discovered  in the  paper
\cite{Itzhaki:2005tu}.


\section{DBI action}\label{second}
The starting point of our analysis is Dirac-Born-Infeld
action for Dp-brane in a general background
\begin{eqnarray}\label{actgen}
S=-\tau_p\int d^{p+1}x
e^{-\Phi}\sqrt{-\det \bA_{\mu\nu}} \ ,
\bA_{\mu\nu}=\gamma_{\mu\nu}+F_{\mu\nu} \ ,
\nonumber \\
\end{eqnarray}
where $\tau_p$ is Dp-brane tension,
$\Phi(X)$ is dilaton and where
$\gamma_{\mu\nu} \ , \mu,\nu=0,\dots,
p$ is embedding of the metric to the
world volume of Dp-brane
\begin{equation}
\gamma_{\mu\nu}=g_{MN}\partial_\mu X^M
\partial_\nu X^N \ , M,N=0,\dots, 9 \ .
\end{equation}
In (\ref{actgen}) the form $F_{\mu\nu}$
is defined as
\begin{equation}
F_{\mu\nu}=b_{MN}\partial_\mu X^M
\partial_\nu X^N+\partial_\mu A_\nu -
\partial_\nu A_\mu \ ,
\end{equation}
where $b_{MN}$ are components of NS two
form. Thanks to the diffeomorphism
invariance of the world volume theory
it is natural to fix some space time
coordinates $X^M$ to be equal to the
world volume coordinates $x^\mu$:
\begin{equation}
X^\mu=x^{\mu} \ , \mu=0,\dots, p \ .
\end{equation}
Note that this choice generally leads
to the induced metric $\gamma_{\mu\nu}$
and two form $F_{\mu\nu}$ to  be
functions of $x^\mu $ and $X^I$ where
$I,J,...=p+1,\dots, D$ label coordinate
transverse to the world volume of
Dp-brane:
\begin{eqnarray}
\gamma_{\mu\nu}=g_{\mu\nu}(x^\mu, X^I)
+g_{\mu I}(x^\mu,X^I)\partial_\nu X^I+
g_{J\nu}(x^\mu,X^I)\partial_\mu X^J+\nonumber \\
+g_{IJ}(x^\mu,X^I)\partial_\mu X^I
\partial_\nu X^J \ , \nonumber \\
F_{\mu\nu}=
b_{\mu\nu}(x^\mu, X^I)
+b_{\mu I}(x^\mu,X^I)\partial_\nu X^I+
b_{J\nu}(x^\mu,X^I)\partial_\mu X^J+
\nonumber \\
+b_{IJ}(x^\mu,X^I)\partial_\mu X^I
\partial_\nu X^J
+\partial_\mu A_\nu-\partial_\mu A_\nu
\ .\nonumber \\
\end{eqnarray}
In the following we will consider the
case when the background  metric and two
form field are block diagonal so that
$g_{\mu I}=g_{J\nu}=b_{I\nu}=0$. Now the equation
of motion for $X^K \ , K=p+1,\dots,9$
 takes the form
\begin{eqnarray}
\partial_K[e^{-\Phi}]\sqrt{-\det\bA}+
\frac{e^{-\Phi}}{2}\left[
\partial_K g_{\mu\nu}+
\partial_Kg_{IJ}\partial_\mu X^I
\partial_\nu X^J+\right. \nonumber \\
\left.+\partial_Kb_{\mu\nu}+
\partial_Kb_{IJ}\partial_\mu X^I\partial_\nu X^J\right]
\bAi^{\nu\mu}\sqrt{-\det\bA}-\nonumber \\
-\partial_\mu\left[e^{-\Phi}
g_{KI}\partial_\nu X^I\bAi^{\nu\mu}_S
\sqrt{-\det\bA}\right]-
\nonumber \\
-\partial_\mu\left[
e^{-\Phi}
b_{KI}\partial_\nu X^I
\bAi^{\nu\mu}_A\sqrt{-\det\bA}
\right]=0  \ ,
\nonumber \\
\end{eqnarray}
where
\begin{equation}
\bAi^{\mu\nu}_S=\frac{1}{2}
\left(\bAi^{\nu\mu}+\bAi^{\mu\nu}\right) \ ,
\bAi^{\mu\nu}_A=\frac{1}{2}
\left(\bAi^{\nu\mu}-\bAi^{\mu\nu}\right) \ .
\end{equation}
Finally, we should also determine the equation of motion
for the gauge field $A_\mu$:
\begin{equation}
\partial_\nu \left[e^{-\Phi}\bAi^{\nu\mu}_A\sqrt{-\det\bA}
\right]=0 \ .
\end{equation}
The power of the Lagrangian formalism
is that  it is manifestly covariant.
Then it is natural to  determine all
possible conserved charges or currents,
whose divergences are nonzero and that
correspond to some explicitly broken
symmetry in given action. The knowledge
of these   currents then considerably
simplifies the analysis of the dynamics
of Dp-brane. To do this  we will be
more general and start with  the action
\begin{equation}
S=\int d^D x \mL(m^A,
\phi^I(x),\partial_\mu \phi^I(x)) \ ,
\end{equation}
where $m^A \ , A=1,\dots,K$ are
parameters that are  contained in the
action and where $\phi^I$ are dynamical
 fields.
 Let us now  consider the variation
of the action  under general
transformation of coordinates $x^\mu $
in the form
\begin{equation}
x'^{\mu}=x^{\mu}+\delta x^\mu \ ,
\end{equation}
where the infinitesimal $\delta x^\mu$
is specified by a set of infinitesimal
parameters $\omega^i$
\begin{equation}
\delta x^\mu= X^\mu_i(x)\delta \omega^i
\ .
\end{equation}
Under such a transformation the field
$\phi^I$ will, in general, transforms.
Thus
\begin{equation}
\phi'^I(x')=
\phi^I(x)+\delta \phi^I(x) \ ,
\end{equation}
where $\delta \phi^I$ is also
specified by parameters $\delta \omega^i$
so
\begin{equation}
\delta \phi^I=\Phi^I_i(x)\delta \omega^i
\  ,
 \end{equation}
where $\Phi^I$ are functionals of $
\phi^J$ and functions of $x$.
Evidently, the total variation $\delta \phi^I$
derives both from the variation of
the field function and from the variation
of its argument:
\begin{equation}
\phi'^I(x')=\phi'^I(x+\delta x)=
\phi'^I(x)+\partial_\mu \phi^I(x)\delta x^\mu
=
\phi^I(x)+\delta_0\phi^I(x)+\partial_\mu
\phi^I(x)\delta x^\mu \ .
\end{equation}
Thus we find
\begin{equation}
\delta_0\phi^I(x)=
\delta\phi^I(x)-
\partial_\mu \phi^I(x)\delta x^\mu=
[\Phi^I_i-\partial_\mu \phi^I]\delta
\omega^i  \ .
\end{equation}
 Let us also presume that the
action is  invariant under these
symmetry transformations when
parameters $m^A$ transform as well
\begin{equation}
m'^A=m^A+\delta m^A=
m^A+\Omega^A_i(\phi,x)\delta\omega^i \ .
\end{equation}
Then the variation of the action takes
the form
\begin{equation}
\delta S= \int \delta (d^Dx)\mL+ \int
d^Dx\delta \mL
 \ ,
\end{equation}
where $\delta \mL$ is the variation in
the Lagrangian density caused by above
variation of $x^\mu, \phi^I$ and $m^A$
and $\delta \left(d^Dx\right)$ is the
variation of the integration measure
caused by variation of $x^\mu $. In
fact
\begin{eqnarray}
\delta \left(d^Dx\right)= d^Dx'-d^Dx=
\partial_\mu (X^\mu _i \omega^i)d^Dx \
.
\nonumber \\
\end{eqnarray}
Now remember that the variation of the
Lagrangian density is equal to
\begin{equation}
\delta \mL(x)=
\delta_0\mL(x)+\frac{\delta \mL}{\delta m^A}
\delta m^A+\partial_\mu \mL \delta x^\mu \ ,
\end{equation}
where the variation $\delta_0\mL$ is
caused by variation of
$\delta_0\phi^I$. Then the variation of
$S$ is
\begin{eqnarray}
\delta S= \int d^Dx\partial_\mu
\left(\mL X^\mu_i\delta \omega^i+
\frac{\delta \mL}{\delta \partial_\mu
\phi^I} \delta_0\phi^I\right) +\int
d^Dx \frac{\delta \mL}{\delta m^A}
\Omega^A_i\delta\omega^i +\nonumber
\\
+\int d^Dx\left[ \frac{\delta
\mL}{\delta \phi^I} -\partial_\mu
\frac{\delta \mL}{\delta
\partial_\mu \phi^I}\right]\delta_0\phi^I
 \ . \nonumber \\
\end{eqnarray}
If the fields obey the equation of
motion that the expression on the last
line  vanishes. Then for the fields
that obey the equation of motion  the
variation of the action is equal to
\begin{equation}\label{dS}
\delta S=-\int d^Dx
\partial_\mu (j^\mu_i \delta \omega^i)+
\int d^Dx \frac{\delta \mL}{\delta m^A}
\Omega^A_i\delta\omega^i \ ,
\end{equation}
where
\begin{equation}
j^\mu_i=\left(-\mL\delta ^\mu_\nu
+\frac{\delta \mL}{\delta \partial_\mu
\phi^I}\partial_\nu
\phi^I\right)X^\nu_i -\frac{\delta
\mL}{\delta \partial_\mu
\phi^I}\Phi^I_i \ .
\end{equation}
Now suppose that $S$ is invariant when
the variations are parameterized by
constant $\delta \omega^i$. Then
(\ref{dS}) implies
\begin{equation}
0=\int d^Dx\left(-\partial_\mu j_i^\mu
+\frac{\delta \mL}{\delta m^A}
\Omega^A_i\right)\delta \omega^i \ .
\end{equation}
Consequently we get that the  current
$j^\mu_i$ obeys the equation
\begin{equation}\label{dja}
\partial_\mu j^\mu_i=
\frac{\delta \mL}{\delta
m^A}\Omega^A_i \ .
\end{equation}
In case when $\Omega^A_i$ vanish we obtain
familiar result that the  charge
\begin{equation}
Q_i(t)\equiv
\int d^{D-1}x j^0_i
\end{equation}
is constant, independent on $t$.

Now we apply this general discussion to
the case of Dp-brane where as we know,
the  world volume fields are $X^I$ and
$A_\mu$. Considering parameters $m^A$,
we will introduce them into the action
in case when the action is not
invariant without their transformations
as well.

 As the first example we
determine components of the world
volume stress energy tensor. To do this
we consider following transformation
\begin{equation}
x'^\mu=x^\mu+\epsilon^\mu \ ,
\end{equation}
where $\epsilon^\mu=const$.
This transformation
corresponds to
\begin{equation}\label{xmunu}
X^\mu_\nu=\delta^\mu_\nu \ .
\end{equation}
Under this transformation the world
volume fields transform as follows
\begin{eqnarray}\label{str}
X'^I(x')=X^I(x) \Rightarrow
\Phi^{X^I}_\mu=0 \ ,
\nonumber \\
A'_\mu(x')=A_\nu(x)
\left(\frac{\partial x'^\mu}{\partial
x^\nu}\right)^{-1}=
A_\nu(x)\delta^\nu_\mu=A_\mu(x)
\Rightarrow \Phi^{A_\mu}_\nu=0 \ .
\nonumber \\
\end{eqnarray}
Since the action is manifestly
invariant under these transformations
there is no need to presume that the
parameters $m^A$ transform as well  and
the right hand side of the equation
(\ref{dja}) is zero. Now using
(\ref{xmunu}) and (\ref{str}) we obtain
the components of the world volume
stress energy tensor
\begin{eqnarray}\label{stressenergytensor}
j^\mu_\nu\equiv T^\mu_\nu=\left(-\mL\delta^\mu_\kappa
+\frac{\delta \mL}{\delta \partial_\mu X^I}
\partial_\kappa X^I+\frac{\delta \mL}
{\delta \partial_\mu A_\sigma}
\partial_\kappa A_\sigma\right)\delta^\kappa_\nu=
\nonumber \\
=\tau_pe^{-\Phi}
\left(\delta^\mu_\nu-g_{IJ}\partial_\kappa
 X^J\bAi^{\kappa\mu}_S\partial_\nu
 X^I-\right.
 \nonumber \\
 \left. -
b_{IJ}\partial_\kappa X^J\bAi^{\kappa\mu}_A
\partial_\nu X^I
-\partial_\nu
A_\sigma\bAi^{\sigma\mu}_A\right)
\sqrt{-\det\bA} \
\nonumber \\
\end{eqnarray}
that obey the equation
\begin{equation}\label{conT}
\partial_\mu T^\mu_\nu=0 \ .
\end{equation}
 Even if it seems that
the analysis performed above is well
known  we mean  that it was  useful to
do it. Especially the case when the
symmetry is explicitly  broken will be
important in the next section.
\section{D1-brane in the background
of I-brane}\label{third} In this
section we will study the dynamics of
D1-brane in the  background studied in
the work \cite{Itzhaki:2005tu}. Namely,
we consider  the intersection of two
stack of NS5-branes. We have $k_1$
NS5-branes extended in $(0,1,2,3,4,5)$
directions and the set of $k_2$
NS5-branes extended in $(0,1,6,7,8,9)$
directions. Let us define
\begin{eqnarray}
\by=(x^2,x^3,x^4,x^5) \ , \nonumber \\
\bz=(x^6,x^7,x^8,x^9) \ .
\nonumber \\
\end{eqnarray}
We have $k_1$ NS5-branes localized
at the points $\bz_n \ n=1,\dots,k_1$
and $k_2$ NS5-branes localized
at the points $\by_a \ , a=1\dots,k_2$.
Every pairs of fivebranes from
different sets intersect at different
point $(\by_a,\bz_n)$.
The supergravity background corresponding
to this configuration takes
the form
\begin{eqnarray}\label{bg}
\Phi(\bz,\by)=\Phi_1(\bz)+
\Phi_2(\by) \ , \nonumber \\
g_{\mu\nu}=\eta_{\mu\nu} \ ,
\mu,\nu=0,1 \ , \nonumber \\
g_{\alpha\beta}=e^{2(\Phi_2-
\Phi_2(\infty))}\delta_{\alpha\beta} \
, \mathcal{H}_{\alpha\beta\gamma}=
-\epsilon_{\alpha\beta\gamma\delta}
\partial^\delta \Phi_2 \ ,
\alpha,\beta,\gamma,\delta=
2,3,4,5 \ , \nonumber \\
g_{pq}=e^{2(\Phi_1-\Phi_1(\infty))}
\delta_{pq} \ ,
\mathcal{H}_{pqr}=-\epsilon_{pqrs}
\partial^s\Phi_1 \ ,
p,q,r,s=6,7,8,9 \ , \nonumber \\
\end{eqnarray}
where $\Phi$ on the
first line means the dilaton
and where
\begin{eqnarray}
e^{2(\Phi_1-
\Phi_1(\infty))}=1+
\sum_{n=1}^{k_1}
\frac{l_s^2}{|\bz-\bz_n|^2} \ ,
\nonumber \\
e^{2(\Phi_2-
\Phi_2(\infty))}=1+\sum_{a=1}^{k_2}
\frac{l_s^2}{|\by-\by_a|^2} \ .
\nonumber \\
\end{eqnarray}
Our goal is to study properties of this
background from the point of view of
D1-brane probe when $\bz_n=\by_a=0$.
The action for D1-brane moving in given
background takes the form
\begin{equation}
S=-\tau_1\int d^2x
e^{-\Phi}\sqrt{-\det\bA} \ ,
\end{equation}
where we have implicitly
 used the
static gauge so that the matrix $\bA$
is equal to
\begin{equation}
\bA_{\mu\nu}=g_{\mu\nu}+
g_{IJ}\partial_\mu X^I
\partial_\nu X^J+
b_{IJ}\partial_\mu X^I
\partial_\nu X^J+\partial_{\mu}A_\nu-
\partial_\nu A_\mu \ , I,J=2,\dots,9 \ .
\end{equation}
To simplify notation let us  denote
\begin{equation}
e^{2(\Phi_1-\Phi_1(\infty))}=H_1(\bz) \ ,
e^{2(\Phi_2-\Phi_2(\infty))}=H_2(\by)  \ ,
\end{equation}
where for coincident branes we have
\begin{equation}
H_1=1+\frac{k_1l_s^2} {|\bz|^2} \ ,
H_2=1+\frac{k_2l_s^2} {|\by|^2} \ .
\end{equation}
Let us now consider the probe in the
near horizon limit where
\begin{equation}
\frac{k_1l_s^2}
{|\bz|^2}\gg 1 \ ,
\frac{k_2l_s^2}
{|\by|^2}\gg 1 \
\end{equation}
so that  we can write
\begin{equation}
H_1=\frac{\lambda_1}
{|\bz|^2} \ , \lambda_1=k_1l_s^2 \ ,
H_2=\frac{\lambda_2}
{|\by|^2} \ , \lambda_2=k_2l_s^2 \ .
\end{equation}
In the near horizon limit the action
takes the form
\begin{equation}\label{D1nhl}
S=-\tau_1\int d^2x
\frac{1}{g_1g_2\sqrt{H_1H_2}}
\sqrt{-\det \bA} \ ,
\end{equation}
where
\begin{eqnarray}
\bA_{\mu\nu}=
\eta_{\mu\nu}+H_1\delta_{pr}
\partial_\mu Z^p\partial_\nu Z^r+
H_2\delta_{\alpha\beta}\partial_\mu
Y^\alpha\partial_\nu Y^\beta+\nonumber
\\
 B_{pr}\partial_\mu Z^p\partial_\nu
Z^r +B_{\alpha\beta}\partial_\mu
Y^\alpha
\partial_\nu Y^\beta+\partial_{\mu}A_{\nu}
-\partial_\nu A_\mu \ ,
\nonumber \\
\end{eqnarray}
and where $g_1=e^{\Phi_1(\infty)} \ ,
g_2=e^{\Phi_2(\infty)}$. According to
arguments given in the paper
\cite{Itzhaki:2005tu} we can expect
that D1-brane that moves in the near
horizon limit possess some additional
symmetry.  In fact, it is easy to guess
 the form of   such a
 transformation
 \begin{equation}\label{Gtr}
Z'^p(x')=\Gamma Z^p(x) \  ,
Y'^\alpha(x')=\Gamma^{-1}Y^\alpha(x) \
, A'_\mu(x')=A_\mu(x) \ , x'^\mu=x^\mu
\ ,
\end{equation}
where $\Gamma$ is a  real parameter.
Now we will argue that the action (
\ref{D1nhl}) is not generally invariant
under the transformation ( \ref{Gtr}).
Firstly, using (\ref{Gtr}) we
 see that  $H_1$ and $H_2$ transform
as
\begin{eqnarray}\label{Hbar}
H_1(Z')=\frac{\lambda_1}{\Gamma^2
|\bZ|^2}=\Gamma^{-2}H_1(Z) \ ,
\nonumber \\
H_2(Y')=\frac{\lambda_2\Gamma^2}{
|\bY|^2}=\Gamma^2 H_2(Y) \ . \nonumber
\\
\end{eqnarray}
so that $\sqrt{H_1H_2}$ is invariant
under (\ref{Gtr}). On the other hand
using ( \ref{Hbar}) we also see
that $\Phi_1$ and $\Phi_2$ transform as
\begin{equation}
\Phi_1(Z')=-\ln \Gamma +\Phi_1(Z) \
, \Phi_2(Y')=\ln \Gamma +\Phi_2(Y) \
.
\end{equation}
Then  (\ref{bg}) implies  that
$\mathcal{H}_{pqr}(Z)$ and
$\mathcal{H}_{\alpha\beta\gamma}(Y)$
transform as
\begin{equation}
\mathcal{H}_{pqr}(Z')=\Gamma^{-1}H_{pqr}(Z)
\ ,
\mathcal{H}_{\alpha\beta\gamma}(Y')=\Gamma
H_{\alpha\beta\gamma}(Y) \ .
\end{equation}
Finally from the definition
$\mathcal{H}=db$ we get that  $b'_{pq}$
and $b'_{\alpha\beta}$ are  invariant
under the transformations (\ref{Gtr})
\begin{equation}\label{Btrs}
b_{pq}(Z')=  b_{pq}(Z) \  ,
b_{\beta\gamma}(Y')= b_{\beta\gamma}(Y)
\ .
\end{equation}
This result implies that
the expressions $b_{pq}\partial_\mu
Z^p\partial_\nu Z^q \ ,
b_{\alpha\beta}\partial_\mu Y^\alpha
\partial_\nu Y^\beta$ break the
scale invariance of DBI action.
In order to restore it
we  introduce two parameters $m^1,m^2$
as
\begin{equation}
b_{pq}\partial_\mu Z^p\partial_\nu Z^q
\rightarrow m^1b_{pq}\partial_\mu
Z^p\partial_\nu Z^q
\end{equation}
and also
\begin{equation}
b_{\alpha\beta}\partial_\mu Y^\alpha
\partial_\nu Y^\beta
\rightarrow
m^2b_{\alpha\beta}\partial_\mu Y^\alpha
\partial_\nu Y^\beta \ .
\end{equation}
Then using (\ref{Btrs}) we see that in
order to have an action invariant under
this scaling transformation parameters
$m_1$ and $m_2$ have to transform as
\begin{equation}\label{mst}
m'^1=\Gamma^{-2}m^1 \ , m'^2=\Gamma^2
m^2 \ .
\end{equation}
Finally using   (\ref{Gtr}) and
(\ref{mst}) we obtain  following values
of $\Phi^I$ and $\Omega^A$
\begin{eqnarray}\label{gtri}
\Phi^{Z^p}_D=Z^p \ ,
 \Phi^{Y^\alpha}_D=-Y^\alpha \ ,
 \Phi^{A_\mu}_D=0 \ ,
\Omega^{m^1}_D=-2m^1 \ ,
\Omega^{m^2}_D=2m^2
\ . \nonumber \\
\end{eqnarray}
As a final remark  note that the
original action  can be recovered from
the action containing $m_1$ and $m_2$
simply by taking  $m^1=m^2=1$.

Now using (\ref{gtri}) we  get the form
of  the dilatation current
\begin{eqnarray}\label{dilcg}
j^\mu_D=-\frac{\delta \mL}{\delta
\partial_\mu Z^p}\Phi^{Z^p}_D-
\frac{\delta \mL}{\delta \partial_\mu
Y^\alpha}\Phi^{Y^\alpha}_D=\nonumber \\
\tau_1e^{-\Phi}
\left(g_{pr}Z^p\partial_\nu Z^r
\bAi^{\nu\mu}_S+b_{pr}Z^p
\partial_\nu Z^r\bAi^{\nu\mu}_A
\right)\sqrt{-\det\bA}-\nonumber \\
-\tau_1e^{-\Phi}
\left(g_{\alpha\beta}Y^\alpha
\partial_\nu Y^\beta\bAi^{\nu\mu}_S
+b_{\alpha\beta}Y^\alpha
\partial_\nu Y^\beta \bAi^{\nu\mu}_A
\right)\sqrt{-\det\bA} \ .
\nonumber \\
\end{eqnarray}
 Finally, we will calculate the
term that appears   on the right hand
side of the equation (\ref{dja})
\begin{eqnarray}
\frac{\delta \mL}{\delta m^A}
\Omega^A_D= -2\frac{\delta \mL}{\delta
m^1} m^1+2\frac{\delta \mL}{\delta
m^2}m^2=
\nonumber \\
= \tau_1e^{-\Phi}
\left(b_{pq}\partial_\mu Z^p
\partial_\nu Z^q-b_{\alpha\beta}
\partial_\mu Y^\alpha\partial_\nu
Y^\beta\right)\bAi^{\nu\mu}\sqrt{-\det
\bA} \ , \nonumber \\
\end{eqnarray}
where we have taken
$m^1=m^2=1$ in the end of calculation.

From the form of this anomaly term we
see that the conservation of $j^\mu_D$
is restored for homogenous modes or for
modes that are spatial depend only
since then the anomaly term vanishes
thanks to the antisymmetry of $b$. In
this paper we restrict to the case of
time dependent modes only. The case
when world volume fields are spatial
dependent will be discussed in separate
publication.

Before we proceed we would like to
write the expression for Hamiltonian
density of Dp-brane moving in
nontrivial background. The reason why
we do this is that the Hamiltonian for
D1-brane will play an important role in
next section. More details considering
the Hamiltonian density for Dp-brane in
general background can be found, for
example in
\cite{Lindstrom:1997uj,Lindstrom:1999tk}.
In fact, using the analysis given there
we have shown in \cite{Kluson:2005ab}
 that the   Hamiltonian
density for Dp-brane in curved
background takes the form
\begin{eqnarray}\label{hde}
\mH=\sqrt{-g_{00}}
\sqrt{\mK}+\pi^ab_{a0} \ , \nonumber \\
\mK= \Pi_Ig^{IJ}\Pi_J +
\pi^a\gamma_{ab}\pi^b+
d_ag^{ab}d_b+e^{-2\Phi}\tau_p^2
\det\bA_{ab} \  ,
\nonumber \\
\Pi_I=P_I+\pi^ab_{aI}=
P_I+\pi^a(b_{aI}+\partial_a
X^Jb_{JI}) \ ,
  \nonumber \\
d_a=\Pi_I\partial_aX^I+ F_{ab}\pi^b \ ,
\nonumber \\
\gamma_{ab}=g_{ab}+
\partial_a
X^Ig_{IJ}\partial_b X^J \ ,
\bA_{ab}=g_{ab}+\partial_a X^I
\partial_b X^Jg_{IJ}+
F_{ab} \ ,  \nonumber \\
F_{ab}=\partial_a A_b-\partial_b A_a+
b_{ab}+b_{IJ}\partial_aX^I\partial_bX^J
 \ , \nonumber \\
\end{eqnarray}
where $a,b=1,\dots p$ label the spatial
coordinates on the world volume of
Dp-brane. Note that in (\ref{hde}) we
presume the background where the metric
$g$ and two form $b$ are block diagonal
 so that $b_{aI}=g_{aI}=0$.
Moreover, we  will also  consider the
background with $b_{a0}=0$ and
$g_{00}=-1$. Now using (\ref{hde}) it
is easy to see that the canonical
equations of motion for $X^K$ and $A_a$
take the form
\begin{eqnarray}\label{eqXK}
\partial_0 X^I=
\frac{\delta H}{\delta P_I}=
\frac{g^{IJ}\Pi_J+\partial_a
X^Ig^{ab}d_b}{\sqrt{\mK}} \ ,
\nonumber \\
\partial_0 A_a=
\frac{\delta H}{\delta \pi^a}=
\frac{b_{aI}g^{IJ}\Pi_J
+\gamma_{ab}\pi^b
-F_{ab}g^{bc}d_c}{\sqrt{\mK}} \ .
\nonumber \\
\end{eqnarray}
On the other hand the equation of
motion for $P_I$ are much more
complicated. However we will not need
their explicit forms  since we use an
existence of conserved charges. On the
hand the equation of motion for $
\pi^a$ will be useful
\begin{eqnarray}\label{eqpia}
\partial_0 \pi^a=
-\frac{\delta H}{\delta A_a}=
\partial_c\left[\frac{\pi^a g^{cd}d_d
-\pi^cg^{ad}d_d} {\sqrt{\mK}}\right]-
\partial_b\left[\frac{e^{-2\Phi}
\tau_p^2}{\sqrt{\mK}}\bAi^{ba}_A
\det\bA
\right] \ .  \nonumber \\
\end{eqnarray}

\section{Time dependent world volume fields}
\label{fourth} In this section we will
study the  case when  all world volume
modes are time dependent only.
Moreover, for the D1-brane the
situation simplifies further since now
$a,b=1$ and hence the quantities
defined in (\ref{hde}) are equal to
\begin{equation}
\gamma_{11}=g_{11} \ ,
\bA_{11}=g_{11} \ , d_1=0 \ ,
\Pi_I=P_I \ .
\end{equation}
Now thanks to the manifest rotation
symmetry $SO(4)$ in the subspaces
spanned by coordinates
$\bz=(z^6,z^7,z^8,z^9)$ and
$\by=(y^2,y^3,y^4,y^5)$ we can reduce
the problem to the study of the motion
on two dimensional subspaces, namely we
will presume that only following world
volume modes are excited
\begin{equation}
z^6=R \cos \theta \ , z^7=R \sin \theta
\ ,
\end{equation}
and
\begin{equation}
y^2=\tR \cos\psi \ ,
y^3=\tR \sin\psi \ .
\end{equation}
For these modes $\mK$ defined in
(\ref{hde})  is equal to
\begin{eqnarray}\label{Ktd}
\mK= P_Rg^{RR}P_R+P_Sg^{SS}P_S
+P_\theta g^{\theta\theta}P_\theta+
P_\psi g^{\psi\psi}P_\psi+
\pi^1g_{11}\pi^1+
e^{-2\Phi}\tau_1^2g_{11}=
\nonumber \\
=\frac{\tau_1^2 R^2S^2} {g_1^2g_2^2
\lambda_1\lambda_2}
\left(1+\frac{\lambda_2g_1^2g_2^2}{
\tau_1^2S^2}P_R^2 + \frac{\lambda_1
g_1^2g_2^2}{\tau_1^2R^2}P^2_S+
\frac{\lambda_2g_1^2g_2^2}{
\tau_1^2R^2S^2}P_\theta^2+
\frac{\lambda_1
g_1^2g_2^2}{\tau_1^2R^2S^2}P^2_\theta
+\frac{g_1^2g_2^2
\lambda_1\lambda_2}{\tau_1^2 R^2S^2}
\pi^2 \right) \ ,
\nonumber \\
\end{eqnarray}
where on the second line
  we have taken  the near horizon
  limit. Note that  for simplicity of notation
we define $\pi\equiv \pi^1 \ ,
A\equiv A_1$.  Then using (\ref{Ktd})
 the Hamiltonian density
(\ref{hde}) takes the form
\begin{equation}\label{hdetc}
\mH=\frac{\tau_1 RS} {g_1g_2
\sqrt{\lambda_1\lambda_2}}
\sqrt{1+\frac{\lambda_2g_1^2g_2^2}{
\tau_1^2S^2}P_R^2 + \frac{\lambda_1
g_1^2g_2^2}{\tau_1^2R^2}P^2_S+
\frac{\lambda_2g_1^2g_2^2}{
\tau_1^2R^2S^2}P_\theta^2+
\frac{\lambda_1
g_1^2g_2^2}{\tau_1^2R^2S^2}P^2_\theta
+\frac{g_1^2g_2^2
\lambda_1\lambda_2}{\tau_1^2 R^2S^2}
\pi^2 } \ .
\end{equation}
 In the same way we obtain  that the
zero component of the dilatation
current $j^0_D$ (\ref{dilcg}) is equal
to
\begin{equation}\label{dde} j^0_D\equiv
d= -P_RR+P_{\tR}\tR \ .
\end{equation}
Since the resulting densities do not
depend on $x^1$ we will work with  them
instead of  with corresponding charges.

Firstly, it is easy to see, using
(\ref{hdetc}), that
\begin{eqnarray}
\dot{P}_\theta=-\frac{\delta
\mH}{\delta \theta}=0 \ ,
\nonumber \\
\dot{P}_\psi=-\frac{\delta
\mH}{\delta \psi}=0 \ ,
\nonumber \\
\dot{\pi}=-\frac{\delta
\mH}{\delta A}=0 \
\nonumber \\
\end{eqnarray}
so that $P_\psi,P_\theta$ and $\pi$ are
constant. On the other hand  the
equations of motion for $R$ and $\tR$
take the form
\begin{eqnarray}\label{dotr}
\dot{R}=\frac{\delta \mH}{\delta
P_R}=\frac{\tau_1 R\tR}
{g_1g_2\sqrt{\lambda_1\lambda_2}}
\frac{\lambda_2g_1^2g_2^2}{\tau_1^2
\tR^2}\frac{P_R}{\sqrt{(\dots)}}=
\frac{R^2P_R}{\lambda_1\mH} \ ,
\nonumber \\
\dot{\tR}=\frac{\delta \mH}{\delta
P_{\tR}}=\frac{\tau_1 R\tR}
{g_1g_2\sqrt{\lambda_1\lambda_2}}
\frac{\lambda_1g_1^2g_2^2}{\tau_1^2
R^2}\frac{P_{\tR}}{\sqrt{(\dots)}}=
\frac{\tR^2P_{\tR}}{\lambda_2\mH} \ ,
\nonumber \\
\end{eqnarray}
where we have used the fact that
$\mH$ is conserved.
Now in order to find the time
dependence of $R,S$ we proceed in the
standard way and use  an existence of
the conserved dilatation density
(\ref{dde}) and conserved hamiltonian
density (\ref{hdetc}). We begin with
simpler case when
$P_\psi=P_\theta=\pi=0$.
\subsection{Case of
zero $P_\theta,P_\psi,\pi$} In this
case the Hamiltonian density is equal
to
\begin{equation}\label{hdeo}
\mH=\frac{\tau_1R\tR}{
g_1g_2\sqrt{\lambda_1\lambda_2}}
\sqrt{1+
\frac{\lambda_2g_1^2g_2^2}{\tau_1^2
\tR^2} P_R^2+
\frac{\lambda_1g_1^2g_2^2}{\tau_1^2
R^2}P_{\tR}^2} \ .
\end{equation}
Let us also consider the situation when
the dilatation density (\ref{dde})
vanishes and hence
\begin{equation}\label{prf}
\frac{P_R}{\tR}=\frac{P_{\tR}}
{R} \ .
\end{equation}
Inserting  (\ref{dotr}) into
 the equation (\ref{prf})  we get
\begin{equation}
\frac{\dot{R}}{R}\lambda_1
=\frac{\dot{\tR}}{\tR}\lambda_2
\end{equation}
that implies
\begin{equation}\label{rmr}
R^{\frac{\lambda_1}{\lambda_2}}C=\tR \ ,
\end{equation}
where $C=e^{\frac{d_0}{\lambda_2}}$
 is an integration  constant and where the
meaning of $d_0$  will be given below.  With the
help of (\ref{prf}) and (\ref{rmr}) we
can rewrite  the hamiltonian density
(\ref{hdeo}) into the form
\begin{equation}
\mH=\frac{C
R^{(\lambda_1+\lambda_2)/\lambda_2}}
{g_1g_2\sqrt{\lambda_1\lambda_2}}
\sqrt{\tau_1^2+g_1^2g_2^2(\lambda_1+\lambda_2)
C^{-2}P^2_RR^{-2\lambda_1/\lambda_2}}
\end{equation}
that allows us to express $P_R$ as a
function of $\mH$ and $R$
\begin{equation}\label{prrh}
P^2_R=\frac{\mH^2\lambda_1\lambda_2}
{(\lambda_1+\lambda_2)R^2}
-\frac{C^2\tau_1^2}{g_1^2g_2^2(\lambda_1+
\lambda_2)}R^{2\lambda_1/\lambda_2} \ .
\end{equation}
Finally, from (\ref{dotr}) and
(\ref{prrh}) we get a differential
equation for $R$ in the form
\begin{equation}\label{dotR}
\dot{R}= \pm R \sqrt{
\frac{\lambda_2}{\lambda_1
(\lambda_1+\lambda_2)} - \frac{\tau_1^2
R^{\frac{2(\lambda_1+\lambda_2)}{
\lambda_2}}}
{C^2(g_1g_2)^2\lambda_1^2\mH^2(\lambda_1+\lambda_2)}}
\
\end{equation}
that has the solution
\begin{equation}\label{RSt}
R=\left(\frac{
\lambda_1\lambda_2(g_1g_2)^2 \mH^2}
{C^2 \tau_1^2}\right)^{\frac{\lambda_2}
{2(\lambda_1+\lambda_2)}}
\left(\frac{1}{\cosh
\sqrt{\frac{\lambda_1+\lambda_2}
{\lambda_1\lambda_2}}t}\right)^
{\frac{\lambda_2}{\lambda_1+\lambda_2}}
\ ,
\end{equation}
where we have chosen the initial
condition that for $t=0$ D1-brane is in
its turning point where $\dot{R}=0$.
Finally, using (\ref{rmr}) and
(\ref{RSt}) we obtain the time
dependence of $S$
\begin{equation}\label{RSts}
\ S=C\left(\frac{
\lambda_1\lambda_2(g_1g_2)^2 \mH^2} {
C^2\tau_1^2}\right)^{\frac{\lambda_1}
{2(\lambda_1+\lambda_2)}}
\left(\frac{1}{\cosh
\sqrt{\frac{\lambda_1+\lambda_2}
{\lambda_1\lambda_2}}t}\right)^
{\frac{\lambda_1}{\lambda_1+\lambda_2}}
\ .
\end{equation}
The physical picture of the time
evolution of D1-brane is following. The
D1-brane leaves the world volume of
I-brane at $t=-\infty$, reaches its
turning point at $t=0$ and then it
again moves to the world volume of
I-brane. It would be certainly very
interesting to propose the holographic
interpretation of this dynamic
situation as was done in case of
Dp-brane in NS5-brane background in
\cite{Sahakyan:2004cq}.

Let us now consider more general case
when the dilatation current $j^0_D$ is
nonzero and hence
\begin{equation}\label{ddn}
d=-P_RR+P_SS= -\lambda_1
\mH\frac{\dot{R}}{R}+\lambda_2\mH
\frac{\dot{S}}{S} \ ,
\end{equation}
where we have used the equations of
motion (\ref{dotr}). The integration of
the equation (\ref{ddn}) gives
\begin{equation}\label{RSrd}
\frac{d}{\mH}t+d_0=\ln\left(\frac{S^{\lambda_2}}
{R^{\lambda_1}}\right) \ \Rightarrow
S=R^{\lambda_1/\lambda_2}e^{\frac{1}{\lambda_2}
(\frac{d}{\mH}t+d_0)} \ .
\end{equation}
Then using (\ref{ddn}) and (\ref{RSrd})
we can express the Hamiltonian density
(\ref{hdeo}) as
\begin{eqnarray}\label{hdeod}
\mH=\sqrt{\frac{R^2S^2\tau_1^2}{(g_1g_2)^2
(\lambda_1\lambda_2)}+\frac{1}{\lambda_1}
P_RR^2+\frac{1}{\lambda_2}P_SS}=
\nonumber \\
=\sqrt{\frac{R^2S^2
\tau_1^2}{(g_1g_2)^2
(\lambda_1\lambda_2)}+\frac{\lambda_1+\lambda_2}
{\lambda_1\lambda_2}
R^2P_R^2+\frac{d^2}{\lambda_2}+
2P_RR\frac{d}{\lambda_2}} \ .
\nonumber \\
\end{eqnarray}
 From (\ref{hdeod})
we now express $P_R$ as function of
$R,S$
\begin{eqnarray}
P_R=-\frac{d\lambda_1}{(\lambda_1+\lambda_2) R}\pm
\frac{\lambda_1\lambda_2}{(\lambda_1+\lambda_2)R}
\sqrt{\frac{\mH^2(\lambda_1+\lambda_2)}
{\lambda_1\lambda_2}-\frac{d^2}{\lambda_1
\lambda_2}-\frac{R^2S^2\tau_1^2}
{\lambda (g_1g_2)^2\lambda_1\lambda_2}}
\ .
\nonumber \\
\end{eqnarray}
Consequently the equation of motion
for $R$ takes the form
\begin{eqnarray}\label{dotRnd}
\dot{R}=\frac{R^2P_R}{\lambda_1\mH}=
-d\frac{1}{(\lambda_1+\lambda_2)}R \pm
\frac{\lambda_2
R}{(\lambda_1+\lambda_2) \mH}
\sqrt{\frac{\mH^2(\lambda_1+\lambda_2)}
{\lambda_1\lambda_2}-\frac{d^2}{\lambda_1
\lambda_2}-\frac{(\lambda_1+\lambda_2)R^2S^2\tau_1^2}
{(g_1g_2)^2(\lambda_1\lambda_2)^2}} \ .
\nonumber \\
\end{eqnarray}
In order to solve the equation (\ref{dotRnd})
we consider following ansatz
\begin{equation}\label{RC}
R=K(t)
e^{-\frac{d}{(\lambda_1+\lambda_2)\mH}t}
\end{equation}
and insert it  to the equation
(\ref{dotRnd}).  Then we obtain a
differential equation for $K$
\begin{equation}\label{dotC}
\dot{K}=\pm
\frac{K}{\sqrt{\lambda_1+\lambda_2}}
\sqrt{\frac{\lambda_2}{\lambda_1}
-\frac{d^2}{\mH^2(\lambda_1+\lambda_2)}
\frac{\lambda_2}{\lambda_1}-
\frac{\tau_1^2e^{2d_0/\lambda_2}}{\mH^2(g_1g_2)^2
\lambda_1^2}K^{\frac{2(\lambda_1+\lambda_2)}{\lambda_2}}}
\ ,
\end{equation}
where we have used the fact that
for
(\ref{RC}) we have
\begin{equation}
R^2S^2=
R^{\frac{2(\lambda_1+\lambda_2)}{\lambda_2}}
e^{\frac{2}{\lambda_2}(\frac{d}{\mH}t+d_0)}=
K^{\frac{2(\lambda_1+\lambda_2)}{
\lambda_2}}e^{2d_0/\lambda_2} \ .
\end{equation}
The equation (\ref{dotC})
can be straightforwardly
integrated with the result
\begin{eqnarray}
K=\left(\frac{\sqrt{\lambda_1\lambda_2}
\sqrt{1-\frac{d^2}{\mH^2(\lambda_1+\lambda_2)}}
g_1g_2\mH
e^{-\frac{d_0}{\lambda_2}}}{\tau_1}\right)^
{\lambda_2/(\lambda_1+\lambda_2)}
\left(\frac{1}{\cosh\sqrt{\frac{\lambda_1+\lambda_2}{\lambda_1
\lambda_2}}
\sqrt{1-\frac{d^2}{\mH^2(\lambda_1+\lambda_2)}}t}\right)
^{\lambda_2/(\lambda_1+\lambda_2)} \ .
\nonumber \\
\end{eqnarray}
Finally using the result given above
and (\ref{RC}) we obtain the time
dependence of $R$
\begin{eqnarray}\label{Rfull}
R=e^{-\frac{1}{\sqrt{\lambda_1+\lambda_2}}
\left(\frac{1}{\sqrt{\lambda_1+\lambda_2}}
\frac{d}{\mH}t+\frac{d_0}{\sqrt{\lambda_1+\lambda_2}}\right)}
\left(\frac{\sqrt{\lambda_1\lambda_2}
\sqrt{1-\frac{d^2}{\mH^2(\lambda_1+\lambda_2)}}
g_1g_2\mH}{\tau_1}\right)^{\frac{\lambda_2}{
(\lambda_1+\lambda_2)}}\times
\nonumber \\
\times
\left(\frac{1}{\cosh\sqrt{\frac{\lambda_1+\lambda_2}{\lambda_1
\lambda_2}}\sqrt{1-\frac{d^2}{\mH^2
(\lambda_1+\lambda_2)}t}}\right)^{\frac{\lambda_2}
{(\lambda_1+\lambda_2)}}
\ .
\nonumber \\
\end{eqnarray}
The time dependence of $S$  then
follows from (\ref{RSrd}) and from
(\ref{Rfull}).

 An importance of the
solution given above will be seen below
when we compare it with the solution
obtained in the background
 introduced
in \cite{Itzhaki:2005tu}. To see this
let us again write
the Lagrangian for time dependent D1-brane
in the near horizon limit
\begin{equation}\label{actcom}
\mL=-\frac{\tau_1}{g_1g_2}\sqrt{\frac{R}{\lambda_1}}
\sqrt{\frac{S}{
\lambda_2}}
\sqrt{1-\frac{\lambda_1}{R^2}
\dot{R}^2-\frac{\lambda_2}{S^2}\dot{S}^2} \ .
\end{equation}
As the first step let us introduce two
modes $\phi_1$ and $\phi_2$ defined as
\begin{equation}
\sqrt{\lambda_1}\frac{dR}{R}=d\phi_1 \
, \sqrt{\lambda_2}\frac{dS}{S}=d\phi_2
\
\end{equation}
or equivalently
\begin{equation}\label{RSp}
R=e^{\frac{\phi_1}{\sqrt{\lambda_1}} }\
, S=e^{\frac{\phi_2}{\sqrt{\lambda_2}}} \ .
\end{equation}
If we insert (\ref{RSp}) into
(\ref{actcom}) we obtain
\begin{equation}\label{accom1}
\mL=-\frac{\tau_1e^{
(\frac{\phi_1}{\sqrt{\lambda_1}} +
\frac{\phi_2}{\sqrt{\lambda_2}})}
}{\sqrt{\lambda_1\lambda_2}g_1g_2}
\sqrt{1-\dot{\phi}_1^2-\dot{\phi}_2^2}
\ .
\end{equation}
Following \cite{Itzhaki:2005tu} we
 introduce two modes $ \phi,
x^2$ through the relation
\begin{equation}\label{phix2}
Q\phi=\frac{1}{\sqrt{\lambda_1}}\phi_1+
\frac{1}{\sqrt{\lambda_2}}\phi_2 \ ,
Qx^2=\frac{1}{\sqrt{\lambda_2}}\phi_1-
\frac{1}{\sqrt{\lambda_1}}\phi_2 \ ,
\end{equation}
where
\begin{equation}
Q=\frac{1}{\sqrt{\lambda}} \ ,
\frac{1}{\lambda}=\frac{1}{\lambda_1}+
\frac{1}{\lambda_2} \ .
\end{equation}
Note that the inverse transformations
to (\ref{phix2}) take the forms
\begin{eqnarray}\label{phili}
\phi_1=\frac{1}{\sqrt{\lambda_1+\lambda_2}}
\left(\sqrt{\lambda_1}x^2+\sqrt{\lambda_2}\phi
\right) \ , \nonumber \\
\phi_2=\frac{1}{\sqrt{\lambda_1+\lambda_2}}
\left(\sqrt{\lambda_1}\phi-
\sqrt{\lambda_2}x^2\right) \ . \nonumber \\
\end{eqnarray}
Then using (\ref{phili}) we can express  the
Lagrangian (\ref{accom1}) as
\begin{equation}\label{accom2}
\mL=-\frac{\tau_1}{\sqrt{\lambda_1\lambda_2}
g_1g_2}e^{Q\phi}\sqrt{1-(\partial_0 x^2)^2
-(\partial_0 \phi)^2} \ .
\end{equation}
This result demonstrates an enhancement
of symmetry in the near horizon  region
as was previously shown in
\cite{Itzhaki:2005tu}. In fact the form
of the Lagrangian (\ref{accom2})
suggests to interpret $x^2$ as a mode
that describes the location of D1-brane
in additional world volume direction of
$2+1$ dimensional object. Moreover, due
to the fact that the Lagrangian
(\ref{accom2}) does not explicitly
depend on $x^2$ the momentum $P_2$
conjugate to $x^2$
\begin{equation}
P_2=\frac{\delta
\mL}{\delta\partial_0x^2}=
\frac{\tau_1
e^{Q\phi}}{\sqrt{\lambda_1
\lambda_2}g_1g_2}\frac{\partial_0 x^2}
{\sqrt{1-(\partial_0 x^2)^2
-(\partial_0 \phi)^2}} \
\end{equation}
is conserved. We also
see that the Hamiltonian $H$ defined as
\begin{eqnarray}\label{Hk}
H=\partial_0 x^2P_2+\partial_0\phi
P_\phi- \mL=\frac{\tau_1
e^{Q\phi}}{g_1g_2
\sqrt{\lambda_1\lambda_2}}\frac{1}{
\sqrt{1-(\partial_0 x^2)^2- (\partial_0
\phi)^2}}=\nonumber \\
=\sqrt{P^2_2+P^2_\phi+\frac{\tau_1^2}{\lambda_1\lambda_2
(g_1g_2)^2}e^{2Q\phi}} \  \nonumber \\
\end{eqnarray}
is also conserved and equal to the
energy $E$. Now  using (\ref{Hk}) we
express $P_\phi$ as function of $\phi$
so that the  equation of motion
$\dot{\phi}=\frac{P_\phi}{H}$  implies
a differential equation for $\phi$
\begin{equation}
\frac{d\phi}{ \sqrt{1-\frac{\tau_1^2}
{(E^2-P_2^2)(g_1g_2)^2\lambda_1
\lambda_2} e^{2Q\phi}}}=\pm
\sqrt{1-\frac{P^2_2}{E^2}}dt \ .
\end{equation}
that has the solution
\begin{eqnarray}
e^{Q\phi}=\frac{\sqrt{E^2-P^2_2}(g_1g_2)
\sqrt{\lambda_1\lambda_2}}{\tau_1}
\frac{1}{\cosh
Q\sqrt{1-\frac{P^2_2}{E^2}}t}
\nonumber \\
\end{eqnarray}
or equivalently
\begin{equation}\label{phit}
\phi=\ln\left(\frac{\sqrt{E^2-P_2^2}(g_1g_2)
\sqrt{\lambda_1\lambda_2}}{\tau_1}\right)
^{\sqrt{\frac{\lambda_1\lambda_2}
{\lambda_1+\lambda_2}}}+
\ln\left(\frac{1}{\cosh
Q\sqrt{1-\frac{P^2_2}{E^2}}t}\right)^{
\sqrt{\frac{\lambda_1\lambda_2}
{\lambda_1+\lambda_2}}} \ .
\end{equation}
Note that we have chosen the initial
condition corresponding
 to D1-brane that reaches  its turning
point at $t=0$.  Then from the inverse
transformations (\ref{phili}) and the
fact that $x^2=\frac{P_2}{E}t+x^2_0$ we
obtain the time dependence of $\phi_1$
and $\phi_2$
\begin{eqnarray}
\phi_1=\sqrt{\frac{\lambda_1}{\lambda_1+
\lambda_2}}\left(\frac{P_2}{E}t+x^2_0\right)+
\sqrt{\frac{\lambda_2}{\lambda_1+
\lambda_2}}\phi \ , \nonumber \\
\phi_2=-\sqrt{\frac{\lambda_2}
{\lambda_1+\lambda_2}}\left(\frac{P_2}{E}t+x^2_0\right)+
\sqrt{\frac{\lambda_1}{\lambda_1+
\lambda_2}}\phi \ . \nonumber \\
\end{eqnarray}
Using  (\ref{RSp})
and also (\ref{phit}) we finally
obtain
\begin{eqnarray}
R=\exp\left(\sqrt{\frac{1}{\lambda_1+
\lambda_2}}\left(\frac{P_2}{E}t+x^2_0\right)\right)
\left(\frac{\sqrt{E^2-P_2^2}(g_1g_2)^2\lambda_1
\lambda_2}{\tau_1^2}\right)^{\frac{\lambda_2}
{2(\lambda_1+\lambda_2)}}\times
\nonumber \\
\times \left(\frac{1}{ \cosh
Q\sqrt{1-\frac{P^2_2}{E^2}}t}\right)^{\frac{\lambda_2}{\lambda_1+
\lambda_2}} \ , \nonumber \\
S=\exp\left(-\sqrt{\frac{1}
{\lambda_1+\lambda_2}}\left(\frac{P_2}{E}t+x^2_0\right)\right)
\left(\frac{\sqrt{E^2-P^2_2}(g_1g_2)^2\lambda_1
\lambda_2}{\tau_1^2}\right)^{\frac{\lambda_1}
{2(\lambda_1+\lambda_2)}} \times
\nonumber \\
\times \left(\frac{1}{ \cosh
Q\sqrt{1-\frac{P^2_2}{E^2}}t}\right)^{\frac{\lambda_1}{\lambda_1+
\lambda_2}} \ .
\nonumber \\
\end{eqnarray}
If we compare  this solution with
(\ref{Rfull}) we get following
identification
\begin{equation}
P_2=\frac{d}{\sqrt{\lambda_1+\lambda_2}}
\ ,
x^2_0=\frac{d_0}
{\sqrt{\lambda_1+\lambda_2}} \ .
\end{equation}
 In summary, we have
shown that the scaling symmetry that
was found for D1-brane moving in the
original background has clear relation
to the enhancement symmetry presented
in \cite{Itzhaki:2005tu}. This result
also gives a physical interpretation of
the dilatation charge  $d$ that after
rescaling corresponds to the conserved
momentum $P_2$ in the new flat
direction.
\subsection{Case of nonzero $P_\psi,
P_\theta$ and $\pi$} For nonzero
$P_\psi, P_\theta$ and $\pi$ the
Hamiltonian density takes the form
\begin{equation}\label{hden}
\mH=\frac{ RS} {g_1g_2
\sqrt{\lambda_1\lambda_2}}
\sqrt{\tau_1^2+\frac{\lambda_2g_1^2g_2^2}{
S^2}P_R^2 + \frac{\lambda_1
g_1^2g_2^2}{R^2}P^2_S+
\frac{g_1^2g_2^2}{ R^2S^2}G} \ ,
\end{equation}
where we have defined
\begin{equation}
G=\lambda_2P_\theta^2+
\lambda_1P_\psi^2+
\lambda_1\lambda_2\pi^2 \ .
\end{equation}
As usual, the equations of motion for
$R$ and $S$ take the form
\begin{eqnarray}\label{dotRnp}
\dot{R}=\frac{R^2P_R}{\lambda_1
\mH} \ , \nonumber \\
\dot{S}=\frac{S^2P_S}{\lambda_2
\mH} \ . \nonumber \\
\end{eqnarray}
For simplicity we restrict ourselves
 to the   case
of the  vanishing dilatation density
$d=0$ bear in mind that this analysis
can be easily extended to the case of
nonzero $d$ as we have shown in the previous
section.

For $d=0$  we again obtain the relation
between $R$ and $S$ in the form
\begin{equation}\label{RSnp}
R^{\frac{\lambda_1}{\lambda_2}}C
=S \ , C=e^{\frac{d_0}{\lambda_2}} \ .
\end{equation}
Using also the fact that $P_S=\frac{R}{S}P_R$
we express $\mH$ as function of $R$ and $P_R$ only
\begin{eqnarray}\label{hprr}
\mH=\frac{R\tR}{g_1g_2
\sqrt{\lambda_1\lambda_2}}
\sqrt{\tau_1^2+(g_1g_2)^2\left(
\lambda_2+ \lambda_1\right)
\frac{P_{R}^2}{S^2}+\frac{g_1^2g_2^2}{R^2S^2}G}
=\nonumber \\
=\frac{CR^{\frac{\lambda_1+\lambda_2}{\lambda_2}}}
{g_1g_2\sqrt{\lambda_1\lambda_2}}
\sqrt{\tau_1^2+(g_1g_2)^2(\lambda_1+\lambda_2)C^{-2}P^2_R
R^{-2\lambda_1/\lambda_2}+
(g_1g_2)^2GC^{-2}R^{-2(\lambda_1+\lambda_2)/\lambda_2}}
\ .
\nonumber \\
\end{eqnarray}
If we  express $P_R$ from
 (\ref{hprr}) and insert it to
the first equation in (\ref{dotRnp}) we
obtain the differential equation for
$R$
\begin{eqnarray}
\dot{R}=\pm
\left(\frac{1}{\lambda_1^2\mH^2}
\left[\frac{\mH^2\lambda_1\lambda_2}
{(\lambda_1+\lambda_2)}-
\frac{G}{(\lambda_1+\lambda_2)}\right]R^2
-\frac{1}{\lambda_1^2\mH^2}
\frac{C^2\tau_1^2}{(g_1g_2)^2(\lambda_1
+\lambda_2)}R^{4+2\lambda_1/\lambda_2}
\right)^{1/2} \nonumber \\
\end{eqnarray}
that has the solution
\begin{eqnarray}
R=\left(\frac{\lambda_1\lambda_2(g_1g_2)^2
\mH^2} {C^2
\tau_1^2}-\frac{G(g_1g_2)^2}
{C^2\tau_1^2}\right)^{\frac{\lambda_2}
{2(\lambda_1+\lambda_2)}}
\left(\frac{1}{\cosh
\sqrt{\frac{\lambda_1+\lambda_2}
{\lambda_1\lambda_2}-\frac{G(\lambda_1+
\lambda_2)}{\mH^2(\lambda_1\lambda_2)^2}}t}\right)^
{\frac{\lambda_2}{\lambda_1+\lambda_2}}
\ ,
\nonumber \\
S=C\left(\frac{\lambda_1\lambda_2(g_1g_2)^2
\mH^2} {C^2
\tau_1^2}-\frac{G(g_1g_2)^2}
{C^2\tau_1^2}\right)^{\frac{\lambda_1}
{2(\lambda_1+\lambda_2)}}
\left(\frac{1}{\cosh
\sqrt{\frac{\lambda_1+\lambda_2}
{\lambda_1\lambda_2}-\frac{G(\lambda_1+
\lambda_2)}{\mH^2(\lambda_1\lambda_2)^2}}t}\right)^
{\frac{\lambda_1}{\lambda_1+\lambda_2}}
\ , \nonumber \\
\end{eqnarray}
where on the second line we have used
the relation (\ref{RSnp}). We see that
nonzero values of $P_\theta, P_\psi$
and $\pi$ do not significantly change
the resulting dynamics. The same
situation was previously reported in
the papers devoted to the study of the
dynamics of Dp-brane in NS5-brane
background.
\\
\\
{\bf Acknowledgement}

This work
 was supported in part by the Czech Ministry of
Education under Contract No. MSM
0021622409, by INFN, by the MIUR-COFIN
contract 2003-023852, by the EU
contracts MRTN-CT-2004-503369 and
MRTN-CT-2004-512194, by the INTAS
contract 03-516346 and by the NATO
grant PST.CLG.978785.


\end{document}